\documentclass{conm-p-l}

\copyrightinfo{2009}{}

\usepackage{graphicx}

\usepackage{amssymb,amsmath}

\newtheorem{theorem}{Theorem}[section]

\theoremstyle{definition}

\theoremstyle{remark}

\numberwithin{equation}{section}

\newcommand{\e}{\epsilon}

\newcommand{\Dl}{\Delta}

\newcommand{\ra}{\rightarrow}
\newcommand{\al}{\alpha}
\newcommand{\be}{\beta}

\newcommand{\pa}{\partial}

\newcommand{\bv}{\bar{v}}
\newcommand{\bw}{\bar{w}}

\newcommand{\bp}{\bar{p}}

\newcommand{\na}{\nabla}

\newcommand{\tp}{\tilde{p}}

\newcommand{\non}{\nonumber}

\begin{document}

\title[Stability Criteria of 3D Inviscid Shears]
{Stability Criteria of 3D Inviscid Shears}

\author{Y. Charles Li}

\address{Department of Mathematics, University of Missouri, 
Columbia, MO 65211, USA}

\curraddr{}
\email{liyan@missouri.edu}

\thanks{}

\subjclass{Primary 76, 37; Secondary 35}
\date{}

\dedicatory{}


\begin{abstract}
The classical plane Couette flow, plane Poiseuille flow, and pipe Poiseuille flow 
share some universal 3D steady coherent structure in the form of ``streak-roll-critical layer"
\cite{WGW07}  \cite{Wal03}  \cite{Vis09}. As the Reynolds number approaches infinity, 
the steady coherent structure approaches a 3D limiting shear of the form ($U(y,z), 0, 0$) in velocity variables.
All such 3D shears are steady states of the 3D Euler equations. This raises the importance of investigating 
the stability of such inviscid 3D shears in contrast to the classical Rayleigh theory of inviscid 2D shears. 
Several general criteria of stability for such inviscid 3D shears are derived. In the Appendix, an argument 
is given to show that a 2D limiting shear can only be the classical laminar shear.
\end{abstract}

\maketitle

\section{Introduction}

There has been a lot of continuing interest in searching for 3D steady 
solutions (or traveling wave solutions in a different frame) in plane Couette flow, plane 
Poiseuille flow, and 
pipe Poiseuille flow \cite{OP80} \cite{Nag90} \cite{Wal03} \cite{Ker05} 
\cite{WGW07} \cite{Vis09} \cite{GHC08} \cite{Eck08} \cite{Hof04}. There seems to be confirmation of their 
existence in experiments \cite{Hof04}.   Recent numerical studies of \cite{WGW07}  \cite{Wal03}  
\cite{Vis09} reveal
that the so-called lower branch steady states in the plane Couette flow, plane Poiseuille flow, and pipe 
Poiseuille flow share some universal steady coherent structure in the form of ``streak-roll-critical layer". As 
the Reynolds number approaches infinity, 
the steady coherent structure approaches a 3D limiting shear of the form ($U(y,z), 0, 0$) in velocity 
variables. All the 3D shears of this form are steady states of the 3D Euler equations. This raises 
two important questions:
(1). What is the special property of the limiting shear?  (2). What is the nature of stability of  
3D inviscid shears in contrast to the classical Rayleigh theory of 2D inviscid shears?  The first question 
was addressed in \cite{LV09}. It turns out that the limiting shear satisfies a necessary condition:
$\int \Dl U f(U) dy dz = 0$ for any function $f$. We shall address 
the second question in this study. We shall use the channel flow (plane Couette flow and plane 
Poiseuille flow) as the example.

As the Reynolds number decreases from infinity, the limiting 3D shear as a steady state deforms into the 
lower branch steady state; while the 3D shear itself undergoes a slow drifting toward the classical laminar 
shear. In fact, all the shears (3D and 2D) form a stable submanifold of the classical laminar 
shear. These shears can play a fundamental role in the transition to turbulence from the classical laminar 
shear \cite{LL09b}.

\section{Inviscid Channel Flow}

The inviscid channel flow is governed by the 3D Euler equations
\begin{equation}
\pa_t u_i + u_j u_{i,j} = - p_{,i}, \quad u_{i,i} = 0 ; 
\label{Euler}
\end{equation}
where ($u_1,u_2,u_3$) are the three components of the fluid velocity along 
($x,y,z$) directions, and $p$ is the pressure. 
The boundary condition is the so-called slip condition
\begin{equation}
u_2(x, a, z) = 0, \quad u_2(x, b, z) = 0;
\label{bc}
\end{equation}
where $a<b$,  and $u_i \ (i=1,2,3)$ are periodic in $x$ and $z$ directions with periods $\ell_1$ and $\ell_3$.

We start with the steady shear solutions of the 3D Euler equations:
\[
u_1 = U(y,z),\quad u_2 = 0, \quad u_3 = 0, \quad p = p_0 \ (\text{a constant}),
\]
where $U(y,z)$ is periodic in $z$ with period $\ell_3$. Linearize the 3D Euler equations with the notations
\begin{eqnarray*}
& & u_1 = U(y,z) + \left [ e^{ik(x-ct)} u(y,z) + c.c. \right ], \quad u_2 = e^{ik(x-ct)} v(y,z) + c.c.,  \\
& & u_3 =e^{ik(x-ct)}  w(y,z) + c.c., \quad p \ra p_0+\left [ e^{ik(x-ct)} p(y,z) + c.c.  \right ];
\end{eqnarray*}
where $k$ is a real constant and $c$ is a complex constant, we obtain the linearized 3D Euler equations
\begin{eqnarray}
& & i k (U-c) u+ v U_y + w U_z = - ik p, \label{LC1} \\
& & i k (U-c)v  = - p_y , \label{LC2} \\
& & i k (U-c) w = - p_z , \label{LC3} \\   
& & i k u + v_y + w_z = 0 . \label{LC4}
\end{eqnarray}
Two forms of simplified systems can be derived:
\begin{eqnarray}
& & k^2 (U-c)v = \pa_y \left [ (U-c) (v_y +w_z) - (U_y v +U_z w) \right ] , \label{SE1} \\
& & k^2 (U-c)w = \pa_z \left [ (U-c) (v_y +w_z) - (U_y v +U_z w) \right ] , \label{SE2} 
\end{eqnarray}
with boundary condition $v(a,z)=v(b,z)=0$ and $v,w$ are periodic in $z$; and 
\begin{equation}
(U-c)^2 \na \cdot \left [ (U-c)^{-2} \na p \right ] = k^2 p , \label{SE}
\end{equation}
with boundary condition $\pa_y p(a,z)=\pa_y p(b,z)=0$ and $p$ is periodic in $z$. We are not successful in 
utilizing the system (\ref{SE1})-(\ref{SE2}). System (\ref{SE}) turns out to be fruitful. The first result that can be derived 
from system (\ref{SE}) is the Howard semi-circle theorem. 
\begin{theorem} \cite{How61} \cite{Eck63}
The unstable eigenvalues (if exist) lie inside the semi-circle in the complex plane:
\[
\left ( c_r -\frac{M+m}{2} \right )^2 +c_i^2 \leq   \left (  \frac{M-m}{2}  \right )^2 , \ c=c_r +i c_i , \ c_i > 0 ;
\]
where $M = \max_{y,z} U$, and $m = \min_{y,z} U$.
\label{HSC}
\end{theorem} 
\begin{proof}
Multiply (\ref{SE}) with $\bp$, integrate by parts, and split into real and imaginary parts; we obtain that 
\begin{eqnarray}
& & \int_0^{\ell_3} \int_a^b U G dy dz = c_r \int_0^{\ell_3} \int_a^b G dy dz , \label{sc1} \\
& & \int_0^{\ell_3} \int_a^b U^2 G dy dz = (c_r ^2 +c_i^2)\int_0^{\ell_3} \int_a^b G dy dz , \label{sc2}
\end{eqnarray}
where
\[
G = |U-c|^{-4} \left [ |\na p|^2 +k^2 |p|^2 \right ] .
\]
Let 
\[
M = \max_{y,z} U , \quad m = \min_{y,z} U ,
\]
then
\[
\int_0^{\ell_3} \int_a^b (U-m) (M-U) G dy dz \geq 0 .
\]
Expand this inequality and utilize (\ref{sc1})-(\ref{sc2}), we arrive at the semi-circle inequality in the theorem.
\end{proof}
Our next goal is to find a counterpart of the Rayleigh criterion \cite{DR81}. For that goal, we need to introduce the transform
\begin{equation}
\tp = (U-c)^{-1} p,
\label{pt}
\end{equation}
then $\tp$ satisfies 
\begin{equation}
\Dl \tp + \left [ \frac{\Dl U}{U-c} - \frac{2\na U \cdot \na U}{(U-c)^2}\right ] \tp = k^2 \tp ,
\label{tpe}
\end{equation}
with the boundary condition
\begin{equation}
\pa_y \tp + \frac{U_y}{U-c} \tp = 0 , \text{ at } y=a, b.
\label{tpbc}
\end{equation}
It turns out that we can only derive results when $U$ satisfies the constraint
\begin{equation}
U_y  = 0 , \text{ at } y=a, b;
\label{Uc}
\end{equation}
in this case, $\tp$ satisfies the simplified boundary condition
\begin{equation}
\pa_y \tp = 0 , \text{ at } y=a, b.
\label{stpbc}
\end{equation}
\begin{theorem}
For $U(y,z)$ satisfying the constraint $U_y  = 0$, at $y=a, b$; if $U(y,z)$ has an (inviscid) unstable eigenvalue, then
\begin{enumerate}
\item 
\[
\na \cdot \left ( \frac{1}{|U-c|^4} \na U \right ) = 0 ,
\]
at some point ($y_*, z_*$) in the interior of the domain, for some $c$ ($c_i >0$) in the semi-circle 
$\left ( c_r -\frac{M+m}{2} \right )^2 +c_i^2 \leq   \left (  \frac{M-m}{2}  \right )^2 $, 
where $M = \max_{y,z} U$, and $m = \min_{y,z} U$;
\item 
\[
2 \frac{|c|^2 -U^2}{|U-c|^2} |\na U|^2 +U \Dl U > 0  ,
\]
at some point ($y_*, z_*$) in the interior of the domain, for some $c$ ($c_i >0$) in the semi-circle 
$\left ( c_r -\frac{M+m}{2} \right )^2 +c_i^2 \leq   \left (  \frac{M-m}{2}  \right )^2 $, 
where $M = \max_{y,z} U$, and $m = \min_{y,z} U$.
\end{enumerate}
\label{RF}
\end{theorem}
\begin{proof}
Multiply (\ref{tpe}) with $\bar{\tp}$, integrate by parts, and split into real and imaginary parts; we obtain that 
\begin{eqnarray}
& & \int_0^{\ell_3} \int_a^b \left [ \frac{U-c_r}{|U-c|^2}  \Dl U - 2 \frac{(U-c_r)^2-c_i^2}{|U-c|^4} |\na U|^2 \right ] |\tp |^2 dydz> 0, \label{rayc1} \\
& & \int_0^{\ell_3} \int_a^b \left [ \frac{1}{|U-c|^2}  \Dl U - 4 \frac{U-c_r}{|U-c|^4} |\na U|^2 \right ] |\tp |^2 dydz = 0. \label{rayc2}
\end{eqnarray}
Equation (\ref{rayc2}) directly implies the first claim in the theorem. The second claim is along the spirit of the Fjortoft theorem \cite{DR81}. 
Multiply (\ref{rayc2}) by $c_r$ and add (\ref{rayc1}), we obtain the second claim.
\end{proof}
Next we will derive a relation between $c_i$ and $k$. 
\begin{theorem}
For $U(y,z)$ satisfying the constraint $U_y  = 0$, at $y=a, b$; let $A=\max_{y,z}|\na U|$, $B=\max_{y,z}|\Dl U|$; then the
unstable eigenvalue (if exists) and the wave number $k$ satisfy the condition
\begin{eqnarray*}
& & (kc_i)^2 \leq 2A^2 +B c_i , \quad (c_i >0) ,\\
& & \left [\text{equivalently, } \left ( kc_i - \frac{B}{2k} \right )^2 \leq 2A^2 + \left (\frac{B}{2k}\right )^2 \right ] .
\end{eqnarray*}
\label{ck}
\end{theorem}
\begin{proof}
Multiply (\ref{tpe}) with $\bar{\tp}$ and integrate by parts, we obtain that 
\begin{equation}
 \int_0^{\ell_3} \int_a^b \left [ \frac{\Dl U}{U-c} - 2 \frac{|\na U|^2}{(U-c)^2}  \right ] |\tp |^2dydz = 
 \int_0^{\ell_3} \int_a^b \left [ |\na \tp |^2 +k^2 |\tp |^2 \right ] dydz .
 \label{ckr}
 \end{equation}
 The left hand side of (\ref{ckr}) is less than or equal to 
 \[
 \left ( \frac{B}{|c_i|} + 2 \frac{A^2}{|c_i|^2}  \right ) \int_0^{\ell_3} \int_a^b |\tp |^2dydz .
 \]
 The right hand side of (\ref{ckr}) is greater than or equal to 
 \[
k^2 \int_0^{\ell_3} \int_a^b |\tp |^2dydz .
 \]
Thus
\[
\frac{B}{c_i} + 2 \frac{A^2}{c_i^2}   \geq k^2 , \quad (\text{when } c_i >0);
\]
which leads to the claim of the theorem.
\end{proof}
It is obvious that Theorems \ref{RF} and \ref{ck} apply to the 2D shears $U(y)$ too. Theorem \ref{RF}
is not the exact 3D counterpart of the 2D Rayleigh criterion. The exact counterpart seems elusive. 
Next we will derive a variation formula for the unstable eigenvalue. This type of formulas 
was initially derived by Tollmien \cite{Tol35} \cite{Lin45} for 2D shears. They are useful 
in deriving unstable eigenvalues near neutral eigenvalues. For 3D shears in atmosphere problems
\cite{CS62}, specific approximations can make the stability problem very similar to the 2D 
Rayleigh problem. In such a case, a similar variation formula can also be derived to predict 
unstable eigenvalues near neutral eigenvalues \cite{CS62}. In our current case, no 
approximation can be made, and we have a much harder problem. We have to work with the pressure 
variable of which the singularity nature is not clear even for 2D shears. We can derive a formula 
near an unstable eigenvalue, but its limit to a neutral eigenvalue is elusive and finding a 
neutral eigenvalue here is more challenging than finding an unstable eigenvalue in contrast to the 
2D shear problem. Let ($p,c,k$) and ($p_1,c_1,k_1$) be two unstable eigenfunctions (if exist) to 
(\ref{SE}), then
\begin{eqnarray}
\na \cdot \left [ (U-c)^{-2} \na p \right ] = k^2(U-c)^{-2} p , \label{var1} \\
\na \cdot \left [ (U-c_1)^{-2} \na p_1 \right ] = k_1^2(U-c_1)^{-2} p_1 , \label{var2}
\end{eqnarray}
Multiply (\ref{var1}) by $p_1$ and (\ref{var2}) by $p$, integrate and subtract, we get
\begin{eqnarray*}
& & (c_1-c) \int_0^{\ell_3} \int_a^b \frac{c_1+c-2U}{(U-c_1)^2(U-c)^2}\left [ \na p \cdot 
\na p_1 + k^2 p p_1 \right ] dydz \\
&=& (k_1-k)(k_1+k) \int_0^{\ell_3} \int_a^b \frac{pp_1}{(U-c_1)^2} dydz
\end{eqnarray*}
from which we obtain the variation formula
\begin{eqnarray}
\frac{dc}{dk} &=& -k \int_0^{\ell_3} \int_a^b \frac{p^2}{(U-c)^2} dydz \non \\
& &  \left [ 
 \int_0^{\ell_3} \int_a^b \frac{1}{(U-c)^3}\left [ \na p \cdot 
\na p + k^2 p^2\right ] dydz \right ]^{-1} .
\label{varf}
\end{eqnarray}
The merit of this formula is that it does not involve $dp$. This formula is valid at an 
unstable eigenvalue ($c=c_r+ic_i$, $c_i >0$). If the unstable eigenvalue lies on a curve 
$c=c(k)$ that leads to a neutral eigenvalue $c^0=c(k^0)$, $c_i^0=0$ as in the case of a 2D shear,
then by the semi-circle theorem \ref{HSC}, $c^0= U(y_0,z_0)$ for some ($y_0,z_0$). In such a case,
the limit $k \ra k^0$ of (\ref{varf}) is still very attractive even though finding the neutral 
eigenvalue $c^0$ here is more challenging than finding an unstable eigenvalue in contrast to 
the case of 2D shears. On the other hand, the limit seems very singular (even for 2D shears). First 
of all, in the limit $k \ra k^0$, equation (\ref{SE}) is singular, so the limiting eigenfunction $p$ 
will be singular too. The following simple equation 
\[
t^2q'' +\al tq' +\be q = 0 , \quad (\al , \be \text{ constants})
\]
already show a variety of singular solutions near $t=0$. Moreover, the singularities generated 
by ($U-c^0$) in the integrals in (\ref{varf}) add to the challenge. 

Similarly, starting from (\ref{tpe}), (\ref{Uc}) and (\ref{stpbc}), we can derive the following 
\begin{eqnarray*}
& & (c_1-c) \int_0^{\ell_3} \int_a^b \left [  \frac{\Dl U}{(U-c_1)(U-c)} +\frac{2 \na U \cdot \na U
(c_1+c-2U)}{(U-c_1)^2(U-c)^2} \right ]\tp \tp_1 dydz \\
&=& (k_1-k)(k_1+k) \int_0^{\ell_3} \int_a^b \tp \tp_1 dydz
\end{eqnarray*}
from which we obtain the variation formula
\begin{eqnarray}
\frac{dc}{dk} &=& 2k \int_0^{\ell_3} \int_a^b \tp^2 dydz \non \\
& &  \left [ 
 \int_0^{\ell_3} \int_a^b \left [ \frac{\Dl U}{(U-c)^2} -\frac{4 \na U \cdot \na U}{(U-c)^3} \right 
] \tp^2 dydz \right ]^{-1} .
\label{varf2}
\end{eqnarray}

\section{Viscous Channel Flow}

The viscous channel flow is governed by the Navier-Stokes equations
\begin{equation}
\pa_t u_i + u_j u_{i,j} = - p_{,i} +\e u_{i,jj} , \quad u_{i,i} = 0 ; 
\label{NS-Couette}
\end{equation}
where ($u_1,u_2,u_3$) are the three components of the fluid velocity along 
($x,y,z$) directions, $p$ is 
the pressure, and $\e = 1/R$ is the inverse of the Reynolds number $R$. 
The boundary condition is
\begin{equation}
u_1(x, a, z) = \al , \quad u_1(x, b, z) = \be , \quad
u_j(x, a, z) = u_j(x, b, z) = 0, (j=2,3);
\label{BC-Couette}
\end{equation}
where $a<b$, $\al < \be$, and $u_i \ (i=1,2,3)$ are periodic in $x$ and $z$ 
directions with periods $\ell_1$ and $\ell_3$. For the viscous channel flow,
the 3D shears mentioned above are no longer fixed points, instead they 
drift slowly in time (sometimes called quasi-steady solutions):
\[
\left ( e^{\e t \Dl } U(y,z), 0, 0 \right ).
\]
By ignoring the slow drift and pretending they are still fixed points
(or by using artificial body forces to stop the drifting), 
their unstable eigenvalues will lead to transient nonlinear 
growths as shown numerically \cite{LL09}. The corresponding linear Navier-Stokes 
operator at ($U(y,z), 0, 0$) is given by the following counterpart of 
(\ref{LC1})-(\ref{LC4}), 
\begin{eqnarray}
& & i k (U-c) u+ v U_y + w U_z = - ik p +\e [\Dl -k^2 ] u , \label{vLC1} \\
& & i k (U-c)v  = - p_y +\e [\Dl -k^2 ] v , \label{vLC2} \\
& & i k (U-c) w = - p_z +\e [\Dl -k^2 ] w , \label{vLC3} \\   
& & i k u + v_y + w_z = 0 . \label{vLC4}
\end{eqnarray}
Again two forms of simplified systems can be derived:
\begin{eqnarray}
& & \pa_y \left \{ \left [ \Dl - k^2 -ikR (U-c) \right ] (v_y+w_z) +ikR (
vU_y +w U_z) \right \} \non \\
& & - k^2 (\Dl - k^2) v +ik^3R (U-c) v = 0 , \label{LNS1} \\
& & \pa_z \left \{ \left [ \Dl - k^2 -ikR (U-c) \right ] (v_y+w_z) +ikR (
vU_y +w U_z) \right \} \non \\
& & - k^2 (\Dl - k^2) w +ik^3R (U-c) w = 0 , \label{LNS2}
\end{eqnarray}
with boundary condition 
\[
v(a,z)=v(b,z)=v_y(a,z)=v_y(b,z)=w(a,z)=w(b,z)=0,
\]
and $v,w$ are periodic in $z$; and the other form
\begin{eqnarray}
& & (U-c)^2 \na \cdot \left \{ (U-c)^{-1} [\e (\Dl - k^2)-ik(U-c)]^{-1} \na p \right \} = ikp \non \\
& & +\frac{\e}{ik}(\Dl - k^2)\na \cdot \left \{ [\e (\Dl - k^2)-ik(U-c)]^{-1} \na p \right \}   , \label{PLNS}
\end{eqnarray}
the boundary condition of which is complicated. 

The system (\ref{LNS1})-(\ref{LNS2}) looks quite convenient for numerical 
simulations. Multiply (\ref{LNS1}) by $\bv$ and 
(\ref{LNS2}) by $\bw$, integrate and add the two equations, we obtain the 
following expression for the eigenvalue in term of the eigenfunction:
\[
c_i = \frac{-1}{kRD} [A+E+H+kR \text{ Re}(G) ] , \quad 
c_r = \frac{1}{D} [B + \text{ Im}(G) ] ,
\]
where
\begin{eqnarray*}
A &=& \int \left [ |\na v_y|^2 + |\na w_z|^2 +k^2 |v_y|^2 +k^2 |w_z|^2\right ] \geq 0 , \\
B &=& \int U \bigg (|v_y|^2 + |w_z|^2 +\bv_y w_z + v_y \bw_z \\
& & + k^2 |v|^2 +k^2 |w|^2\bigg ) , \text{ real};\\
D &=& \int \bigg ( | v_y|^2 + | w_z|^2 +\bv_y w_z + v_y \bw_z \\
& & + k^2 |v|^2 +k^2 |w|^2\bigg ) \geq 0 , \\
E &=& \int \left [\bv_{yy} w_{zy} + \bv_{yz} w_{zz} +k^2\bv_y w_z + c.c. \right ] , \text{ real};\\
G &=& -i \int (\bv_y + \bw_z) (vU_y +w U_z) , \text{ complex};\\
H &=& k^2 \int \left [ |\na v|^2 + |\na w|^2\right ] + k^4 \int \left [ |v|^2 + |w|^2\right ] \geq 0 .
\end{eqnarray*}
Notice that the expression of $c_r$ has no explicit dependence upon the 
Reynolds numbber $R$, but it does depends on $R$ implicitly via the eigenfunction. An unstable eigenvalue corresponds to $kc_i >0$. Without loss of generality, 
assume $k >0$. Notice that $A+E \geq 0$, and 
\[
|G| \leq g \int [ |v_y|^2 + |w_z|^2 + |v|^2 + |w|^2] ,
\]
where
\begin{equation}
g = \max \{ \| U_y\|_{L^\infty}, \| U_z\|_{L^\infty} \} . 
\label{gex} 
\end{equation}
Then from the expression of $c_i$, one can obtain the following theorem by comparing the terms $H$ and 
$kR \text{ Re}(G)$.
\begin{theorem}
Let $g$ be given by (\ref{gex}), ($k,c$) be an eigenmode; if $R g < \min ( k, k^3)$, ($k>0$); then $c_i <0$, i.e. a stable eigenvalue.
\label{STB}
\end{theorem}
The type of claims in Theorem \ref{STB} and their improvements have been investigated intensively for 2D 
shears \cite{Syn38} \cite{Jos68} \cite{Jos69}. On the other hand, as mentioned at the beginning, we are 
more interested in unstable eigenvalues and for large Reynolds number $R$ as in \cite{LL09b}. 

\section{Appendix: 2D Limiting Shear Can Only Be the Classical Laminar Shear}

We take the plane Couette flow as the example, for plane/pipe Poiseuille flow, the argument is the same. 
Assume that as the Reynolds number $R \ra +\infty$, a 2D steady state of the plane Couette flow approaches the 
limiting shear ($U(y),0$). Denote by
\[
( U(y) + u(x,y), v(x,y))
\]
the steady state which is periodic in $x$. Assume that ($u,v$) and their spatial derivatives are of order 
$o(R^{-1/2})$ as $R \ra +\infty$. Then to the leading order $O(R^{-1})$,

\begin{eqnarray*}
& & U\pa_x u +v \pa_y U = - \pa_x p +\frac{1}{R} \pa_y^2 U, \\
& & U\pa_x v = - \pa_y p , \\
& & \pa_x u + \pa_y v = 0 .
\end{eqnarray*}
Taking an average in $x-$direction (over the period), we get 
\begin{eqnarray*}
& & \overline{v} \pa_y U = \frac{1}{R} \pa_y^2 U, \\
& & 0= - \pa_y \overline{p} , \\
& & \pa_y \overline{v} = 0 .
\end{eqnarray*}
By the boundary condition of $v$ in $y$ direction, $\overline{v} = 0$. Thus
\[
\pa_y^2 U = 0, \quad \text{i.e. } U = c_1 + c_2 y.
\]
That is, $U$ has to be the laminar linear shear. 

As shown in \cite{LV09}, the corresponding 3D limiting shear $U(y,z)$ does not have to be the  
linear shear, rather satisfies a constraint
\[
\int \Dl U f(U) \ dydz = 0
\]
for any $f$. In fact, the lower branch 3D limiting shear is far away from the linear shear 
\cite{WGW07} \cite{Wal03} \cite{Vis09}. Does this hint that the lower branch steady state does not 
exist in 2D? Numerical simulations could not find any 2D steady state other than the linear shear
\cite{ENR08}.

\end{document}